# Data intensive physics analysis in Azure cloud

*Igor Sfiligoi, Frank Würthwein and Diego Davila
University of California San Diego, La Jolla, CA, USA
*isfiligoi@sdsc.edu, fkw,didavila@ucsd.edu

* Corresponding Author

**Abstract.** The Compact Muon Solenoid (CMS) experiment at the Large Hadron Collider (LHC) is one of the largest data producers in the scientific world, with standard data products centrally produced, and then used by often competing teams within the collaboration. This work is focused on how a local institution, University of California San Diego (UCSD), partnered with the Open Science Grid (OSG) to use Azure cloud resources to augment its available computing to accelerate time to results for multiple analyses pursued by a small group of collaborators. The OSG is a federated infrastructure allowing many independent resource providers to serve many independent user communities in a transparent manner. Historically the resources would come from various research institutions, spanning small universities to large HPC centers, based on either community needs or grant allocations, so adding commercial clouds as resource providers is a natural evolution. The OSG technology allows for easy integration of cloud resources, but the data-intensive nature of CMS compute jobs required the deployment of additional data caching infrastructure to ensure high efficiency.

**Keywords:** Cloud, OSG, data caching, physics.

## 1 Introduction

Scientific computing needs typically vary with time and most fields occasionally experience significant spikes in demand, e.g., right before major conferences. Since provisioning dedicated local resources for peak demand is prohibitively expensive, multi-domain research platforms like the Open Science Grid (OSG) [1] have seen significant success in both aggregating resources located at participating institutions and by provisioning additional resources to scientists in times of need, by either borrowing compute capacity from unrelated science domains or by abstracting access to specialized resources, like XSEDE HPC centers [2]. Adding commercial clouds as resource providers is a natural evolution of the trend.

A major problem when aggregating resources from multiple providers is data handling. While network throughput has improved significantly over time, thanks to projects like the Pacific Research Platform [3], the network latency of wide-area networks is a fundamental physics property and cannot be significantly improved upon. Data-intensive applications that process a variety of non-contiguous remote data can thus incur a significant penalty in such environments, if accessed directly. Many OSG user communities thus rely on caching services, creating an effective content delivery network (CDN) for the most often used data.

This paper is structured as follows. Section 2 provides a short description of the Compact Muon Solenoid (CMS) [4] science community that was the focus of this work. Section 3 provides an overview of the OSG infrastructure. Section 4 provides the technical setup used to integrate Azure cloud resources into OSG, including the CDN services needed by CMS. Finally, section 5 provides an overview of the operational experience running jobs on this setup.

## 2 The CMS collaboration

The Compact Muon Solenoid (CMS) experiment at the Large Hadron Collider (LHC) is operated by a collaboration of more than 200 institutions in more than 40 countries. The experiment produced several exabytes (EB) of data in the first decade of its operations, with over one EB of data currently stored in archival storage alone. In support for all that data, the CMS collaboration operates about 200 PB of active storage infrastructure, globally.

Most of the computing is contributed in-kind by institutions and national funding agencies the world over. The experiment is very versatile and allows many independent high-energy physics (HEP) studies. The data produced during the first decade of operations allowed the scientists to publish more than 1,000 scientific papers. Consequently, at any point in time hundreds of unique analyses are being worked on by the thousands of members of the collaboration. Standard data products are centrally produced, and then used by often competing teams within the collaboration.

This work is focused on how a local institution, University of California San Diego (UCSD), used commercial cloud to augment its available computing to accelerate time to results for multiple analyses pursued by a small group of collaborators. A half dozen collaborators from UCSD, UC Santa Barbara (UCSB), Boston University, and Baylor University accomplished in a few days what would normally take them multiple weeks. This in turn motivated them to pursue additional studies that they would have normally not dared to do, given their compute intensive nature. Cloud integration thus both accelerated their science and allowed them to pursue science that would have otherwise been out of reach.

Some of the analyses that used the cloud resources are summarily described in the subsections below.

## 2.1 The scouting dimuon analysis

The CMS scouting dimuon analysis uses scouting tier data (collision data collected at a higher rate and with smaller event content, compared to normal triggers) to look for displaced dimuon resonances at low masses (as low as twice the muon mass).

Based on the observed and predicted event counts at different dimuon masses, the computing resources are used to calculate the upper limits on various models of new physics, through high dimensional maximum likelihood fits. The fit for one model hypothesis typically takes a few CPU-days, and in this analysis, there were tens of thousands of hypotheses to test. The outcome of the analysis [5] was the exclusion of several models of potentially new physics, e.g., where the Standard Model Higgs boson decays into a pair of hypothetical "dark photons", which can travel some distance in the detector before decaying into a pair of muons.

## 2.2 The two gauge boson analysis

This analysis is looking for a Higgs boson produced in association with two gauge bosons. In the standard model, this is a very rare signature, but deviations from the predicted quartic couplings of two Higgs and two gauge bosons would lead to potentially large excesses at high energies in this final state. In addition, this final state is also sensitive to triple Higgs couplings. As these couplings have never been measured, looking for non-standard values is a high profile endeavor at the LHC. In essence, the Higgs boson that was only discovered in 2012 (and lead to the 2013 Nobel Prize in Physics) is now being used for studies of its properties in order to verify that the Higgs found in nature is indeed the Higgs predicted by theory.

In CMS simulation of event generators, restriction on the phase-space is often required due to the branching fraction of boson decays to reduce the total computation time. The UCSD team performed the full chain of the CMS simulation of event generations for the desired rare processes with an inclusive phase-space. The about 50 M events generated will be used to study and design several future physics analyses leading to potentially multiple publications.

## 2.3 The Top-W scattering analysis

The Top-W scattering analysis searched for a process where a top quark and a W boson scatter on each other, leading to a top quark pair, a W boson plus a high momentum forward quark in the final state. This particular analysis looked for the process in the final state of two leptons of the same charge, a b-tagged jet, and multiple other jets.

Jets are remnants the detector sees as a signature for quarks and gluons. A b-tagged jet is indicative of a b-quark. The top quark decays to a b-quark and a W. A collision with two tops and one W will thus have three W’s and two b-quarks in the final state. This work has thus a lot in common with looking at a multiple-car traffic accident, trying to figure out what happened in the original collision by studying the debris. What was the original accident, and who piled on afterwards.

The targeted process is a sensitive probe of modifications in the couplings between top quarks and the W, Z and Higgs bosons.

## 3 The Open Science Grid Computing Model

The OSG computing model is based on federation principles. There is no central policy entity. Each resource provider is autonomous, both in terms of resource acquisition and user community access policies. Similarly, each user community, also known as a Virtual Organization (VO), is autonomous, both in terms of setting priorities among their constituents and in contracting for access with any resource provider. OSG provides the necessary trust and technical mechanisms, a.k.a. the glue, that allows seamless integration of the many resource providers and user communities without combinatorial issues.

From a technical point of view, OSG provides 4+1 main service categories:

1. An authentication framework, currently based on the Grid Community Toolkit (GCT) X.509 Proxy certificate infrastructure [6].
2. A portal implementation for accessing compute resources at various resource providers, also known as a Compute Entrypoint (CE), alongside a set of standard packages expected on all compute resources.
3. A set of data handling tools and services, including community-specific software distribution mechanism, data storage portals and content-distribution services.
4. A central accounting system [7].
5. An overlay pilot workload management system, i.e. glideinWMS [8]. While not a requirement, most OSG user communities do use it, including the CMS collaboration.

Most of the OSG elements are completely generic and will concurrently serve many user communities. Each individual resource provider decides which VOs to serve, advertising the supported VOs through a central OSG registry [9]. With few exceptions, authentication and authorization is based on group membership, i.e. the resource providers are only concerned with which VO a credential belongs to, and any intra-VO policies are dealt outside the OSG security framework.

Like most OSG communities, the CMS collaboration relies on HTCondor for all its workload management needs, with a significant fraction of resources provisioned through OSG services. The provisioning of compute resources from OSG is regulated by a glideinWMS Fronted, which monitors the job queues and decides whether more resources are needed. When appropriate, it instructs the glideinWMS Factory, operated by OSG as a central service, to provision additional compute resources through one or more Compute Entrypoints using delegated credentials. The Factory provisioning request includes the necessary setup details needed for the provisioned compute resources to securely join the CMS HTCondor pool. From that point on, those resources can be scheduled to run any suitable job; this mode of operation is usually referred to as an overlay or pilot setup. To better understand the system utilization, any job running on OSG resources is recorded in the central OSG accounting system. A schematic overview is available in Fig. 1.

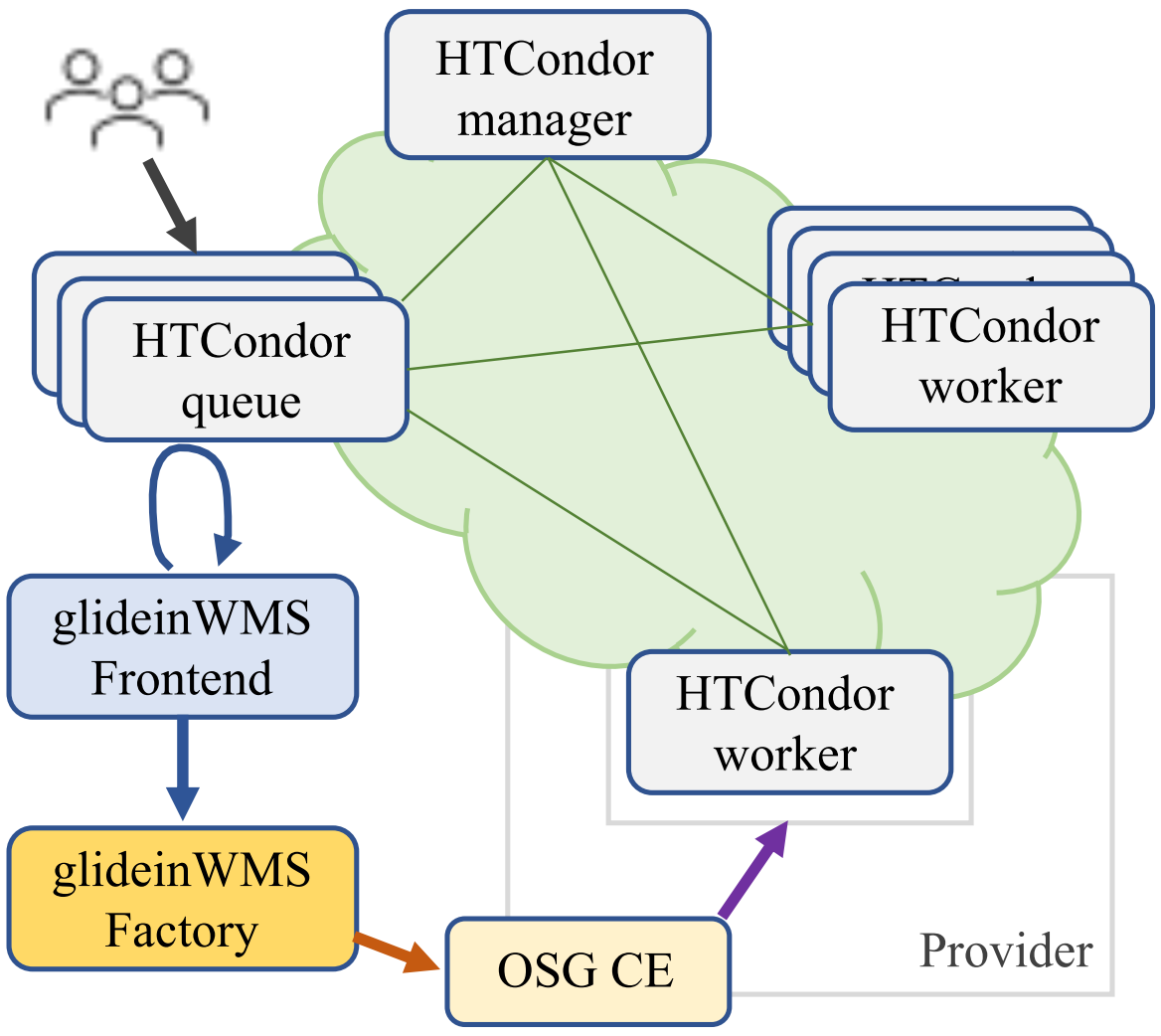

**Fig. 1.** Overview of the OSG workload management system

Note that each CE converts OSG requests into local ones, leaving full control of the provisioning logic to the resource provider. The serviced provisioning requests are however still recorded in the central OSG accounting system. More details on CE internals are available in section 4.

### 3.1 Data handling considerations

CMS jobs rely on CVMFS [10], a POSIX-like read-only distributed file system for software distribution. The CVMFS driver is expected to be installed on all provisioned resources, while all the content is dynamically downloaded by the driver from a central location using the HTTP protocol. Given the relatively static nature of the software, caching servers, based on Apache Squid, should be deployed by resource providers to both compensate for large network latencies and minimize network traffic on the central servers.

Most jobs also require access to the detector calibration parameters, which are relatively big but change very slowly. Many independent jobs will thus access the same data, so CMS relies on an Apache Squid-based content delivery network (CDN), named Frontier [11], to both reduce the access latency and decrease the network bandwidth used. This same infrastructure is typically used for CVMFS caching, too.

Handling of physics data is managed by the jobs themselves, with the bulk of the data being transferred from and to CMS-managed servers. Since this data is contiguous and rarely reused, no OSG-operated support services are needed.

## 4 Azure Cloud Resource Integration

In OSG all resource provisioning details are abstracted behind a portal, a.k.a. the Compute Entrypoint (CE). The used portal implementation is HTCondor-CE [12] and relies on a batch system paradigm, translating global requests into local ones. After successful authentication, the presented credential is mapped to a local system account and all further authorization and policy management is handled in the local account domain, typically at the VO level. Several backend batch systems are supported; we used HTCondor for this work, due to its proven scalability and extreme flexibility.

We used a single HTCondor pool setup per CE to manage resources from many Azure regions. Since HTCondor can easily manage resources distributed over many WAN links and can deal with NAT-ed network environments, we preferred to keep the setup as simple as possible. All connections were secured with mutual authentication using a shared secret, with integrity checks on all transfers. All the worker node resources obviously were provisioned from the Azure cloud. Since OSG computing model allows for preemption, we only used the cheaper preemptible instances [13], also known as spot instances.

We created a custom virtual machine image, starting with the base operating system, then adding the prescribed OSG worker node software, including the CVMFS drivers, and the fully configured HTCondor daemons pointing to our CE pool. Since the images are private, we embedded the shared secret, too. To transparently use the created image in all the targeted cloud regions, we registered the image in the Azure Shared image gallery and used its replication mechanisms.

For ease of use, we used the native Azure interface for resource provisioning, relying on its group provisioning mechanisms, namely the Virtual Machine Scale Sets (VMSS). One VMSS was needed per targeted region. Since only the resource manager is exposed to this interface, we find that the ease of use and flexibility of native interfaces significantly outweighs the added burden of dealing with multiple independent interfaces. With VMSS, one only has to set the desired number of instances at runtime, and each VMSS will provision as much as it is available at that point in time without any further operator intervention. The number of resources requested in each of the regions was mostly manually determined based on the desired spend rate and observed preemption rates. We would prefer less expensive but stable regions at smaller scales but would expand into more expensive regions when observing non-trivial preemption rates in the more favorable ones.

Once provisioned, the worker node instances would join the HTCondor pool and be available to run any jobs present in the queue. HTCondor would manage the priorities between jobs belonging to any number of CE users, just like it would at any HTCondor-managed resource provider. As already mentioned in the previous section, an OSG accounting probe installed on the CE collects usage statistics and registers them with the central OSG accounting system.

To minimize both network latency experienced by the jobs and network volume flowing into the cloud, we also provisioned, installed, and operated one Frontier cache server per employed virtual private network, which in practice meant one instance per Azure cloud region. Note that a Frontier cache sever is really just a properly configured Apache Squid instance. We ensured that the relative IP address was always the same, e.g. X.Y.0.4, making it easy for the worker instances to programmatically determine its location at startup time. Since many worker instances relied on Frontier for its operation, we provisioned them using the more expensive but reliable on-demand instances.

A summary overview of the setup is available in Fig. 2. Note that the depicted HTCondor processes are those managing the internal CE resource pool. The service jobs submitted by the glideinWMS will bring in, configure and start another set of HTCondor processes, as shown in Fig. 1; HTCondor allows for nested resource management.

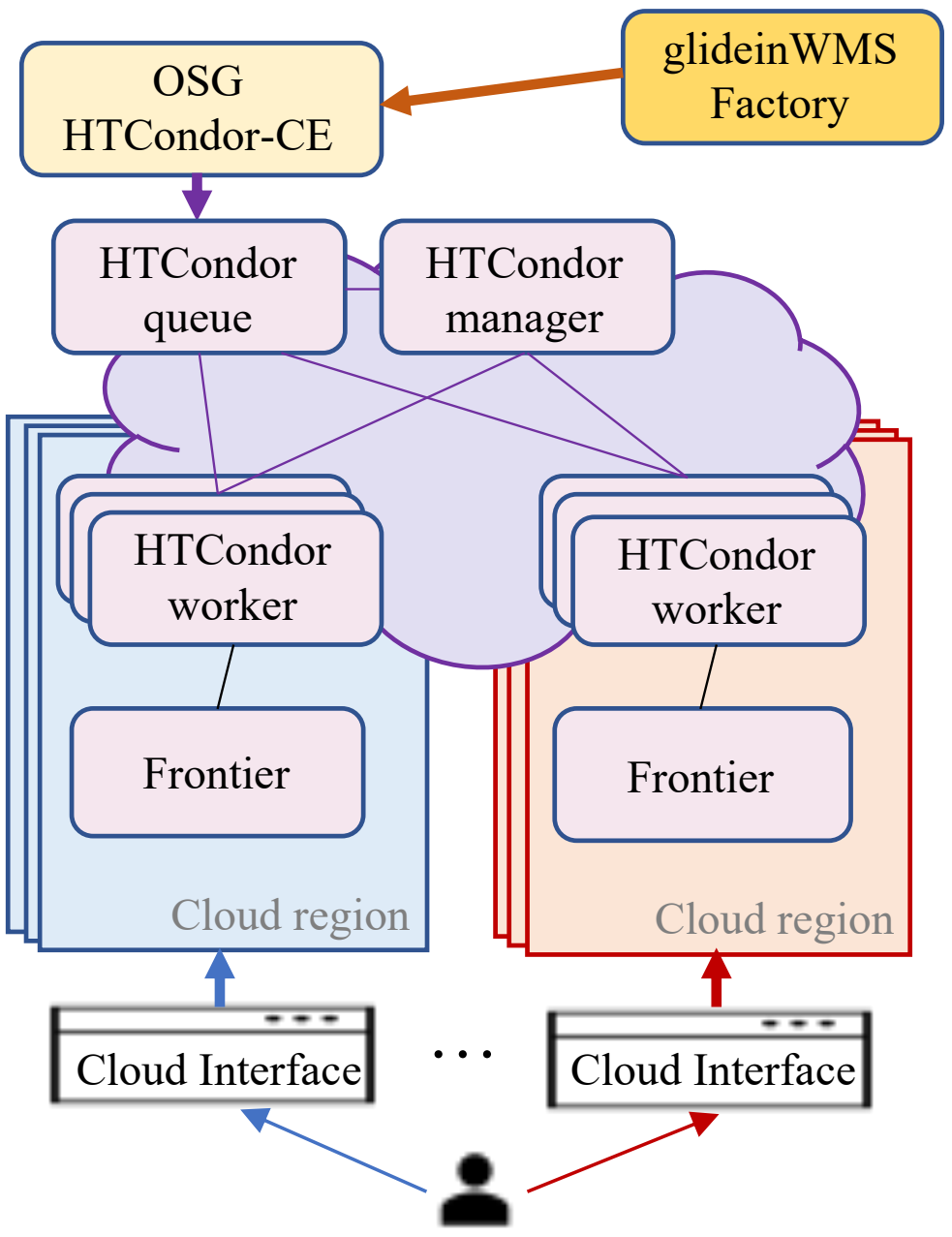

**Fig. 2.** Overview of a cloud-enabled OSG CE

## 5 The operational experience

For convenience, the OSG CE was deployed as an on-demand instance in one of the Azure cloud regions. The public IP was associated with a UCSD-managed DNS record, and then registered with the OSG production services.

CMS jobs are tuned to expect 2 GB of RAM for each CPU core, and prefer utilizing 8 cores concurrently, although a subset of them is single-threaded. At the time of the exercise, i.e. May 2021, the most cost-effective Azure instance type that fit those needs was the F16s v2, providing 16 Intel Xeon Platinum 8272CL CPU cores and 32 GB of RAM at an average cost of $3.3/day per spot instance. HTCondor was configured to dynamically partition the available CPU cores among the users, based on their requests. In order to keep preemption rates reasonably low, we provisioned compute from 5 regions, four in north America and one in Europe. One Frontier cache instance was deployed in each of those regions, too.

As mentioned in section 2, CMS has several user groups who may compete for the same resources. For the purpose of this work, we admitted only pilot jobs serving the UCSD community and pilots serving the Fermilab community, with the UCSD pilots having priority over the Fermilab pilots. We used this setup due to the limited number of users served by the UCSD glideinWMS Frontend, thus resulting in occasionally very spiky usage pattern. Having low-priority jobs in the local HTCondor queue allowed us to make provisioning changes only a couple times a day, while still both providing a high number of CPU cores to the high priority users and fully utilizing the provisioned resources at all times. That said, we limited the UCSD pilots to approximately 20k CPU cores, so the Fermilab pilots did get a significant fraction of the resources when we decided to provision up to 30k CPU cores for a few days.

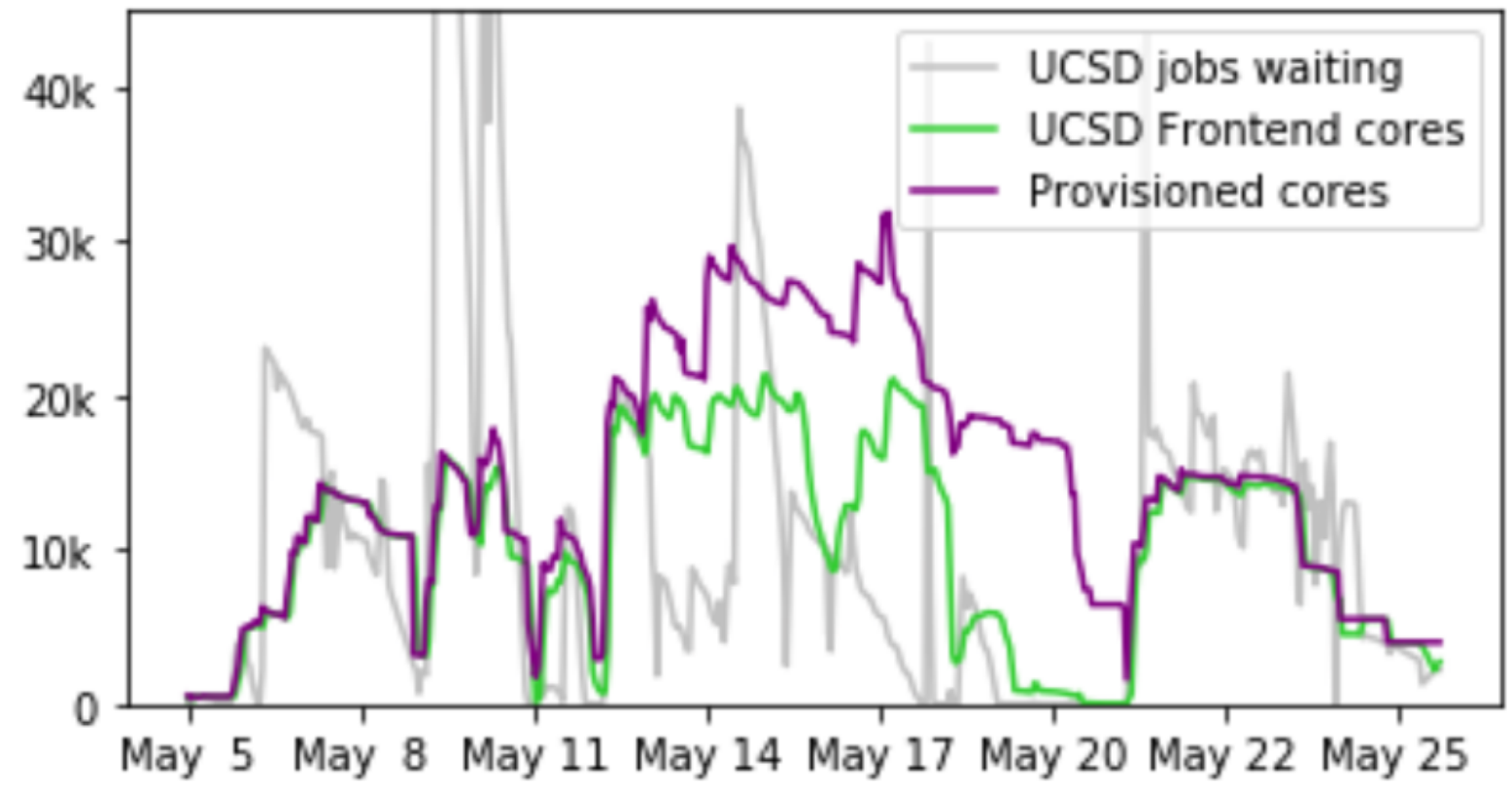

**Fig. 3.** Overview of the provisioned cores and the share going toward UCSD users.

An overview of the provisioned resources and its correlation to the UCSD Fronted is shown in Fig. 3. The figure also shows the impact of Azure preemption at high instance counts; most of the gradual downward slopes are due to Azure preemption alone; we chose not to automatically replace the preempted instances. Note that at lower provisioning levels, e.g. around May 22nd, the line is instead quite flat, indicating that we experienced very little preemption in Azure.

The resource utilization of the compute instances was generally reasonably high. As can be seen from the monitoring snapshot in Fig. 4, the CPU utilization hovered around 80% most of the time, with short-lived dips due to either temporary lack of demand or a change in user jobs' mixture, which introduces inefficacies due to the Frontier cache not having the necessary data in its cache yet. As mentioned before, the UCSD user job activity was rather spiky. A similar utilization pattern would have been expected from an on-prem system, too.

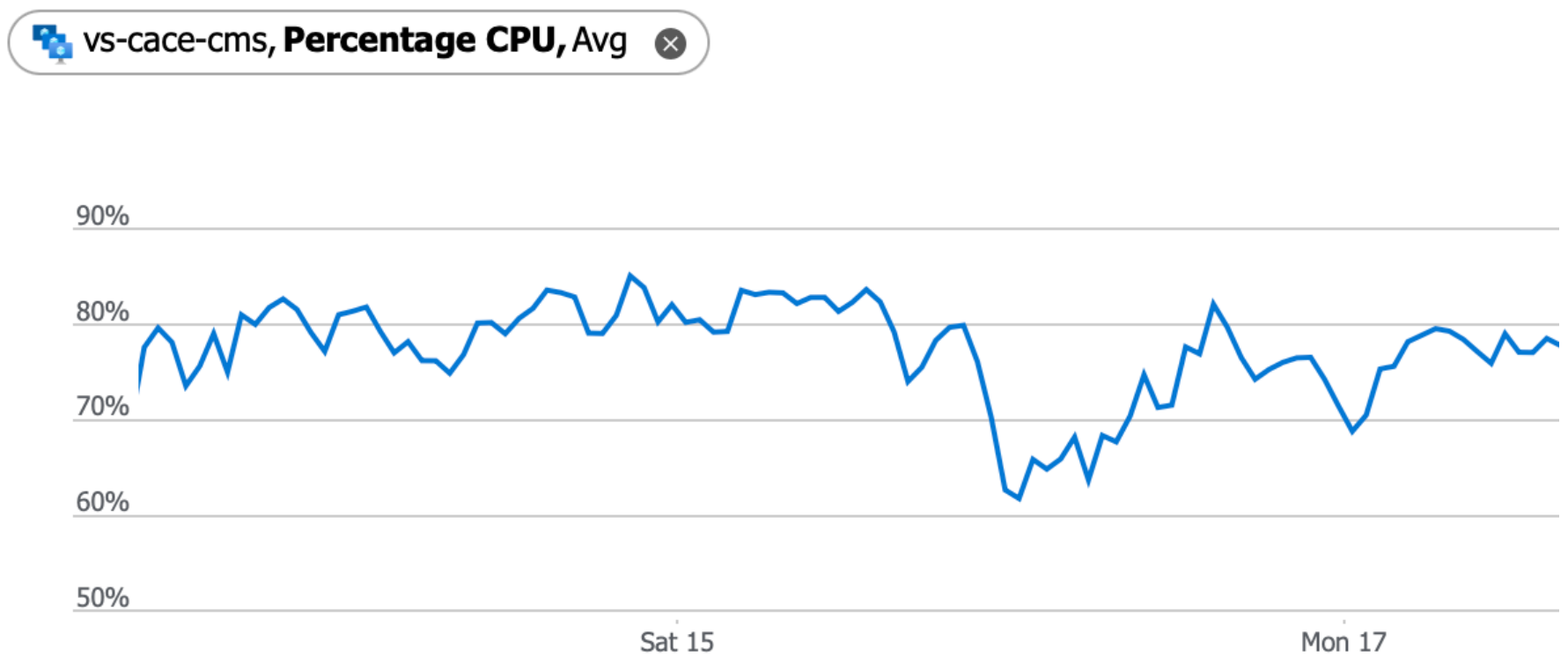

**Fig. 4.** Screenshot of Azure monitoring, showing CPU utilization of one VMSS

The cloud run achieved its goal of doubling the CPU core hours delivered by UCSD for the month of May, using about $70k worth of Azure credits. CMS UCSD on-prem resources typically deliver around 7M core hours each month, and the cloud-enabled CE added another 7.3M. A screenshot of the weekly OSG accounting for the UCSD-associated CMS resources is available in Fig. 5.

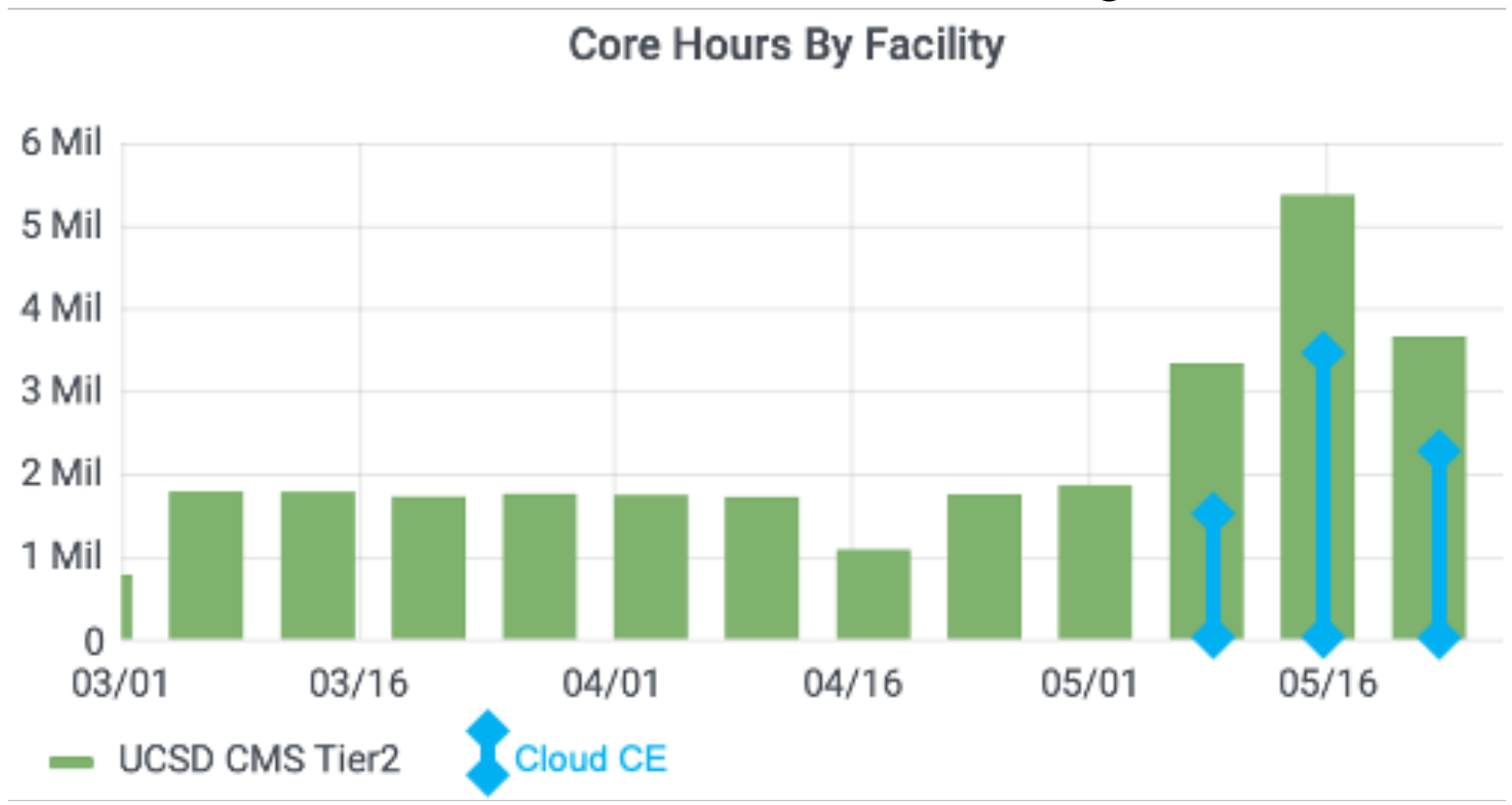

**Fig. 5.** Screenshot of the weekly OSG accounting for UCSD-associated CMS resources.

The additional resources had a very large impact on the produced science, especially because most of them were used to support just a few select analyses of interest to UCSD.

The changes to the user workflow were minimal, too. As an example, the dimuon analysis group reports that they only had to add one HTCondor ClassAd attribute to their job config files, and that alone allowed them to transparently run on the cloud resources. The additional resources allowed them to get an estimated 5x faster turnaround time, i.e., from months to just over a week, compared to using just regular CMS resources as usual, both because of higher peak resource availability, i.e., 20k CPU cores vs more typical 9k, and larger integrated resource availability over the longer period of time. Due to high resource demand and CMS fair share policies, a single CMS user cannot really expect more than O(100k) CPU core hours per week using only the regular CMS resources.

## 6 Summary and conclusion

During the month of May 2021, we more than doubled the compute capacity of the UCSD CMS center by expanding it into the Azure cloud. This was done in such a way that it mostly benefited a set of high profile CMS data analyses the UCSD physics department is involved with. Graduate students, post-docs, and faculty accessed Azure cloud resources with a simple one-line statement in their workload configuration files. The added resources allowed those users to accomplish in days what would have otherwise taken them months to do. Seeing this acceleration in time to science results, several of them added in additional studies they would have normally not dared to do, given their resource intensiveness.

This was possible due to CMS using the OSG provisioning services. By abstracting the cloud resources behind an OSG CE, those resources were essentially identical to the on-prem compute resources those users normally have access to.

Given the data-intensive nature of most CMS analyses, the remote nature of cloud resources required the deployment of content delivery network services in Azure to minimize data-access related inefficiencies. This was particularly urgent given the use of multiple cloud regions, spanning both the US and European locations. The utilized Frontier caches performed greatly, keeping the CPU utilization on par with on-prem resources.

**Acknowledgements.** This work has been partially funded by the US National Science Foundation (NSF) Grants OAC-2030508, MPS-1148698, OAC-1826967, OAC-1836650, OAC-1541349, PHY-1624356, and CNS-1925001. We gratefully acknowledge the credits provided by Microsoft that covered all the Azure cloud expenses.